\documentclass[review]{elsarticle}
\pdfoutput=1
\usepackage[T1]{fontenc}    % use 8-bit T1 fonts
\usepackage{hyperref}       % hyperlinks
\usepackage{url}            % simple URL typesetting
\usepackage{booktabs}       % professional-quality tables
\usepackage{amsfonts}       % blackboard math symbols
\usepackage{nicefrac}       % compact symbols for 1/2, etc.
\usepackage{microtype}      % microtypography
\usepackage{graphicx}
\usepackage{subfigure}
\usepackage{amsmath}
\usepackage{multirow}
\usepackage{tabularx}
\usepackage{array}

\begin{document}

\begin{frontmatter}

\title{Accurate and Rapid Diagnosis of COVID-19 Pneumonia with Batch Effect Removal of Chest CT-Scans and Interpretable Artificial Intelligence}
\author[aimed,dml]{Rassa Ghavami Modegh}
\author[aimed,dml]{Mehrab Hamidi}
\author[aimed]{Saeed Masoudian} 
\author[aimed]{Amir Mohseni}
\author[aimed]{Hamzeh Lotfalinezhad} 
\author[tehran]{Mohammad Ali Kazemi} 
\author[tehran]{Behnaz Moradi} 
\author[iran]{Mahyar Ghafoori} 
\author[iran]{Alireza Aziz-Ahari} 
\author[iran]{Omid Motamedi} 
\author[iran]{Omid Pournik} 
\author[iran2]{Kiara Rezaei-Kalantari} 
\author[isfahan,isfahan2]{Amirreza Manteghinezhad} 
\author[isfahan,isfahan2]{Shaghayegh Haghjooy Javanmard} 
\author[kerman]{Fateme Abdoli Nezhad} 
\author[kerman]{Ahmad Enhesari} 
\author[shahrekord]{Mohammad Saeed Kheyrkhah} 
\author[kashan]{Razieh Eghtesadi} 
\author[kashan2]{Javid Azadbakht} 
\author[kashan]{Akbar Aliasgharzadeh} 
\author[kashan3]{Mohammad Reza Sharif} 
\author[qazvin]{Ali Khaleghi} 
\author[shahidbeheshti]{Abbas Foroutan} 
\author[tehran]{Hossein Ghanaati} 
\author[aimed]{Hamed Dashti} 
\author[aimed,dml]{Hamid R. Rabiee\corref{mycorrespondingauthor}}

\cortext[mycorrespondingauthor]{Corresponding author}
\ead{rabiee@sharif.edu}
%\address{Department of Computer Science and Technology, Ocean University of China}




\address[aimed]{AI-Med Group, AI Innovation Center, Sharif University of Technology, Tehran, Iran}

\address[dml]{DML, Department of Computer Engineering, Sharif University of Technology, Tehran, Iran}

\address[tehran]{Department of Radiology, Tehran University of Medical Sciences, Tehran, Iran}

\address[iran]{Preventive Medicine and Public Health Research Center, Psychosocial Health Research Institute, Community and Family Medicine Department, School of Medicine, Iran University of Medical Sciences, Tehran, Iran}

\address[iran2]{Cardiovascular Medical \& Research Center, Iran University of Medical Sciences, Tehran, Iran}

\address[isfahan]{Applied Physiology Research Center, Isfahan Cardiovascular Research Institute, Isfahan, Iran}

\address[isfahan2]{University of Medical Science, Isfahan, Iran}

\address[kerman]{Department of Radiology, Afzalipour Faculty of Medicine, Kerman University of Medical Sciences, Kerman, Iran}

\address[shahrekord]{Research Institute of Animal Embryo Technology, Shahrekord University, Shahrekord, Iran}

\address[kashan]{Kashan University of Medical Sciences, Kashan, Iran}

\address[kashan2]{Department of Radiology, Kashan University of Medical Sciences, Kashan, Iran}

\address[kashan3]{Department of Pediatrics, Kashan University of Medical Sciences, Kashan, Iran}

\address[qazvin]{Department of Computer Engineering, Imam Khomeini International University, Qazvin, Iran}

\address[shahidbeheshti]{Shahid Beheshti University of Medical Sciences, Medical Academy of Science, Tehran, Iran}


\begin{abstract}
COVID-19 is a virus with high transmission rate that demands rapid identification of the infected patients to reduce the spread of the disease. The current gold-standard test, Reverse-Transcription Polymerase Chain Reaction (RT-PCR), has a high rate of false negatives. Diagnosing from CT-scan images as a more accurate alternative has the challenge of distinguishing COVID-19 from other pneumonia diseases. Artificial intelligence can help radiologists and physicians to accelerate the process of diagnosis, increase its accuracy, and measure the severity of the disease. We designed a new interpretable deep neural network to distinguish healthy people, patients with COVID-19, and patients with other pneumonia diseases from axial lung CT-scan images. Our model also detects the infected areas and calculates the percentage of the infected lung volume. We first preprocessed the images to eliminate the batch effects of different devices, and then adopted a weakly supervised method to train the model without having any tags for the infected parts. We trained and evaluated the model on a large dataset of 3359 samples from 6 different medical centers. The model reached sensitivities of 97.75\% and 98.15\%, and specificities of 87\% and 81.03\% in separating healthy people from the diseased and COVID-19 from other diseases, respectively. It also demonstrated similar performance for 1435 samples from 6 different medical centers which proves its generalizability. The performance of the model on a large diverse dataset, its generalizability, and interpretability makes it suitable to be used as a reliable diagnostic system.
\end{abstract}


\end{frontmatter}

%\linenumbers

\section{Introduction}
\label{sec:introduction}

The novel coronavirus was first identified in China towards the end of 2019, and the World Health Organization (WHO) referred to the virus as COVID-19 \cite{who}. The virus has a high rate of transmission \cite{li2020early} which has terrified people across the world. Given that no confirmed treatment or long-term tested vaccine has been developed for COVID-19 thus far, an accurate and fast diagnosis is vital to reduce the speed of transmission \cite{huang2020clinical, zhu2020novel, hui2020continuing}. 
\par
Although the COVID-19 is typically diagnosed using reverse-transcription polymerase chain reaction (RT-PCR) \cite{zu2020coronavirus, chu2020molecular} as the reference standard for diagnosis of positive infection of coronavirus, the test results in many false negatives, and several studies have shown that Computerized Tomography (CT) scan from chest can be used as a more accurate alternative for the diagnosis of COVID-19 with a low rate of false negatives and above 90\% accuracy \cite{ai2020correlation, li2020coronavirus, lei2020ct}. According to RSNA, some of the COVID-19 patients show no sign of infection in their CT-scan during the first two days of infection, and the anomalies of their CT-scans are not visible \cite{simpson2020radiological, shi2020evolution}. Moreover, a high number of patients visit hospitals to be diagnosed when the spread of the virus is at its peak, which overwhelms the limited number of expert medical staff that can attend to the diagnosis of all patients in a timely manner. An Artificial Intelligence (AI) system can help to accelerate the diagnosis process by aiding the prioritization of high risk cases, and to increase the diagnosis speed and accuracy. The distinction between COVID-19 pneumonia and other lung-related diseases is another challenge for inexperienced radiologists and medical staff that reduce the accuracy of the diagnosis and treatment processes. 
\par
Previous studies have demonstrated that the use of deep networks and artificial intelligence methods can enhance how we tackle the challenges related to COVID-19. A study from China, collected 4356 chest CT-scans from 3,322 patients across 6 hospitals. They developed a model named COVNet that generates a probability score for COVID-19, community-acquired pneumonia (CAP), and non-pneumonia samples. They achieved 90\% sensitivity for COVID-19 patients, 87\% for CAP, and 94\% for non-pneumonia \cite{li2020artificial}. In another study, Wang and Wong designed an architecture called COVID-Net and trained it with 13,975 Chest X-Ray (CXR) images instead of CT-scan images which reached 91\%, 94\%, and 95\% sensitivities for COVID-19, non-COVID-19, and normal cases, respectively \cite{wang2020covid}. In an initial study, Gozes et. al. used multiple international datasets, including disease-infected areas in China. They used ResNet50 which was pre-trained on ImageNet and achieved an Area Under the Curve (AUC) of 0.994 with 94\% sensitivity and 98\% specificity. They also used results collected from 56 COVID-19 patients, and 51 non-COVID-19 patients and achieved a sensitivity of 96.4\% and specificity of 98\% \cite{gozes2020rapid}. Shah et. al. used 738 images from public datasets \cite{zhao2020covid} and designed a new architecture named CTnet-10 that achieved an accuracy of 82.1\%. They also used a pre-trained VGG-19 and transfer learning which resulted in 91.78\% accuracy \cite{shah2020diagnosis}. In a different study, Wang et al. trained a DNN on 453 images of pathogen-confirmed COVID-19 cases. They achieved a total accuracy of 73\% with a specificity of 67\% and a sensitivity of 74\% \cite{wang2020deep}. Xu et al. used 618 CT samples to train a location attention model that separates influenza-A viral pneumonia, COVID-19, and locations that are irrelevant to infection. This location attention model yielded an overall accuracy of 79.4\% \cite{xu2020deep}. Wang et al. used 499 CT scans to train a three-dimensional custom architecture named DeCovNet to distinguish COVID-19 patients from healthy people. They used the labels assigned by the radiologists by investigating the CT-scan images and the clinical symptoms of the patients in training the model. They finally combined the set of regions effective in deciding on the COVID-19 label extracted by the CAM algorithm \cite{zhou2016learning}, and the largest region having a high variance, based on the standard deviation and the number of connected components, in the unsupervised lung segmentation algorithm \cite{liao2019evaluate} to find the one most probable infected region in the sample \cite{wang2020weakly}.
\par 
Despite the success of recent deep learning methods in producing accurate diagnosis of COVID-19 using CT-scan images with high sensitivity, they lack interpretability. So many of them have tested their model on a limited number of samples from one center only which arises the question of whether they have tested their model on different stages of the disease and different imaging conditions. They also do not explicitly address the batch effect removal issues, and thus do not generalize well on images from different sources. Moreover, they mostly utilize supervised machine learning algorithms for training and detection, while supervised methods require a great effort to tag all slices of a large number of CT-scan images. Due to these shortcomings, we propose a new accurate and rapid diagnosis assistant system by utilizing statistical and deep machine learning methods in AI to distinguish patients with COVID-19 from healthy patients and the ones that had diseases other than COVID-19 in an interpretable way, while removing the batch effects. Moreover, the system we designed can highlight the involved regions and calculate the percentage and location of the infected lung volume. We trained our system on a big cohort of CT-scan images collected from different hospitals across different cities in Iran, in order to support CT-scan images with different characteristics.

\section{Method}

\subsection{Workflow}

\begin{figure}[!htb]
\centerline{\includegraphics[width=\columnwidth]{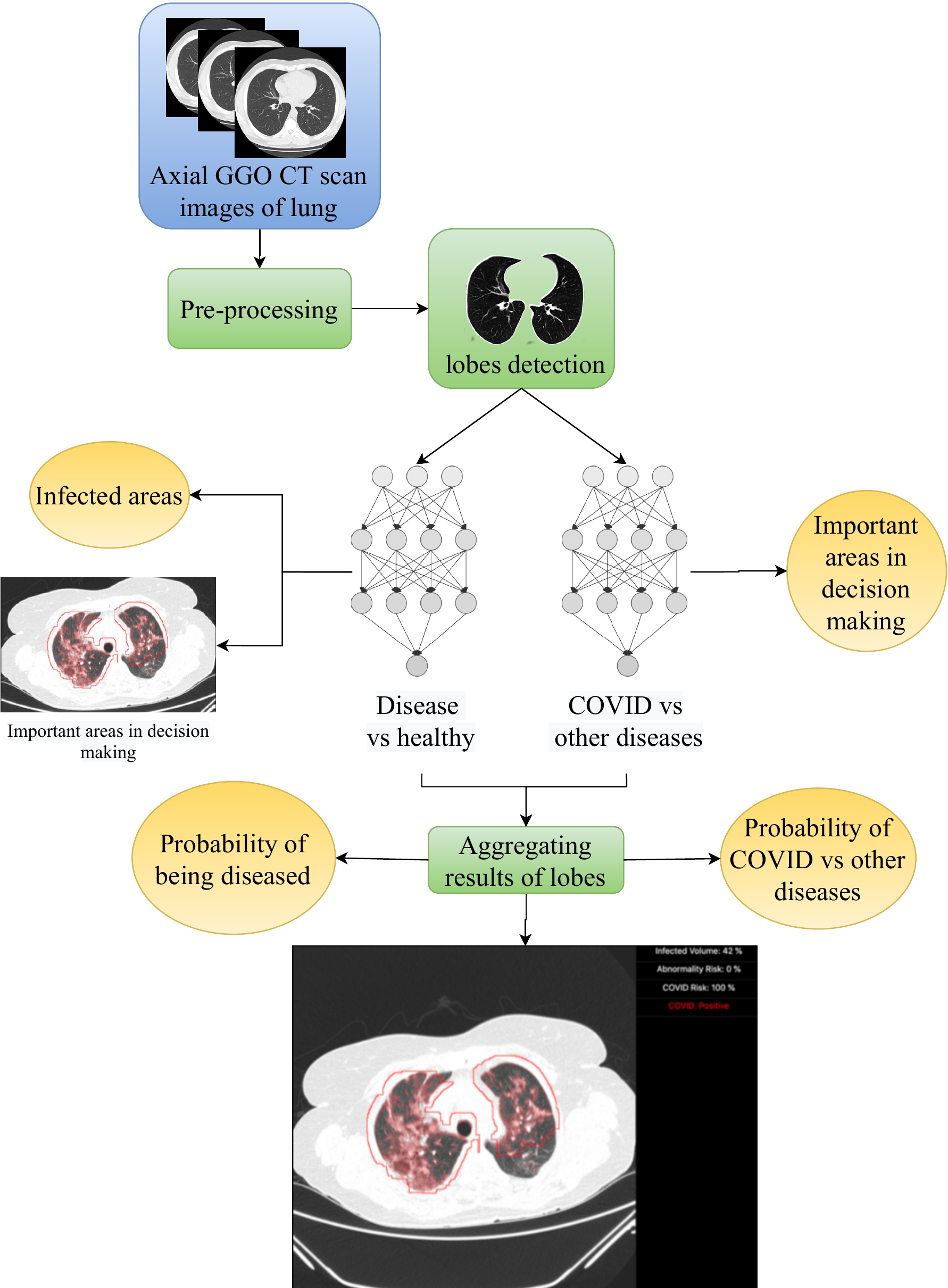}}
\caption{The general workflow for interpretable COVID-19 detection.}
\label{fig1}
\end{figure}

The general workflow for the proposed interpretable COVID-19 detection is shown in Fig. \ref{fig1}. In the first step of pipeline, the Ground Glass Opacity Axial (GGOA) CT-scan images are preprocessed and the lobes of lungs are detected and extracted from the axial slices. The images of the left and right lobes of all the slices are then fed into two deep Convolutional Neural Networks (CNNs), one for calculating the probability of being diseased versus healthy, and the other for calculating the probability of diagnosis to be COVID-19 versus other diseases. In addition to the probability, the first network also detects the infected areas in lung images. In addition, both networks specify the effective areas of the lungs that the decision is made based on them. In the final step, the probabilities assigned to the lobes are aggregated to make a final decision for the whole sample. Each part of the pipeline is explained in detail in the preceding subsections. 

\subsection{Preprocessing of CT-scan images}
The GGOA CT-scan images are stored in DICOM files. These files contain the value calculated by the CT-scan device for each voxel and also the scanner-dependent information for converting the scanned values into the Hounsfield unit. The transformation which is presented in (\ref{eq1}) is linear and requires the values for slope and intercept that are stored in the header of each DICOM file. The values in the Hounsfield unit should range between -2000 to 3000. Since the values out of this range correspond to artifacts, the values lower than -2000 were set to -2000, and the values above 3000 were set to 3000. Thereafter, the minimum value was subtracted from the image to make all the values positive. There were also differences in the background of the images captured by different CT-scan devices. To make all images normalized into the range [0, 2048], the median value of the background in all slices of one CT-scan image was calculated and subtracted from all the values (the negative values were set to 0). The values above 2048 were then clipped into 2048, as they were related to the details in the bones which are irrelevant to the objective of this study. All the values were transformed into a range of [0, 1] by dividing them by 2048. In the final step, a median filter with a kernel size of 3 was applied to eliminate nonlinear noises from the images. 

\begin{equation}
value_{Hounsfield} = slope \times value_{raw} + intercept
\label{eq1}
\end{equation}

\subsection{Detecting lobes}
Each lung image has two similar groups of pixels, those related to soft tissues and bones, and those related to the lung and air. These two groups were separated by applying the K-means clustering algorithm to all pixels in each image. The pixels assigned to the group with the lower value of the center were labeled as the foreground, as the lung and air fall in this group. In the next step, the foreground was eroded by a square kernel of size 3 to eliminate the small noisy isolated areas. The remaining foreground was then dilated by a square kernel of size 8, so the disjoint close areas become connected. Each group of connected pixels was then separated as a single object. The objects with an area less than 5000 pixels, which correspond to small holes, were eliminated. Apart from the background of the lung which expands everywhere, the larger object on the left side of the image was selected as the left lobe, and the larger object on the right side of the image was selected as the right lobe. The whole image inside a rectangular bounding box around each lobe was cropped and copied into the center part of a 256 by 256 image with a black background. In most cases, the cropped image was smaller than the mentioned size. In exceptional cases where the cropped image was bigger than 256 in at least one of the dimensions, it was scaled to fit in a 256 by 256 image for keeping the width to height ratio and then copied into the center of the black background.

\subsection{Infected versus healthy decision maker}
\begin{figure*}[!htb]
\centerline{\includegraphics[width=1.5\columnwidth]{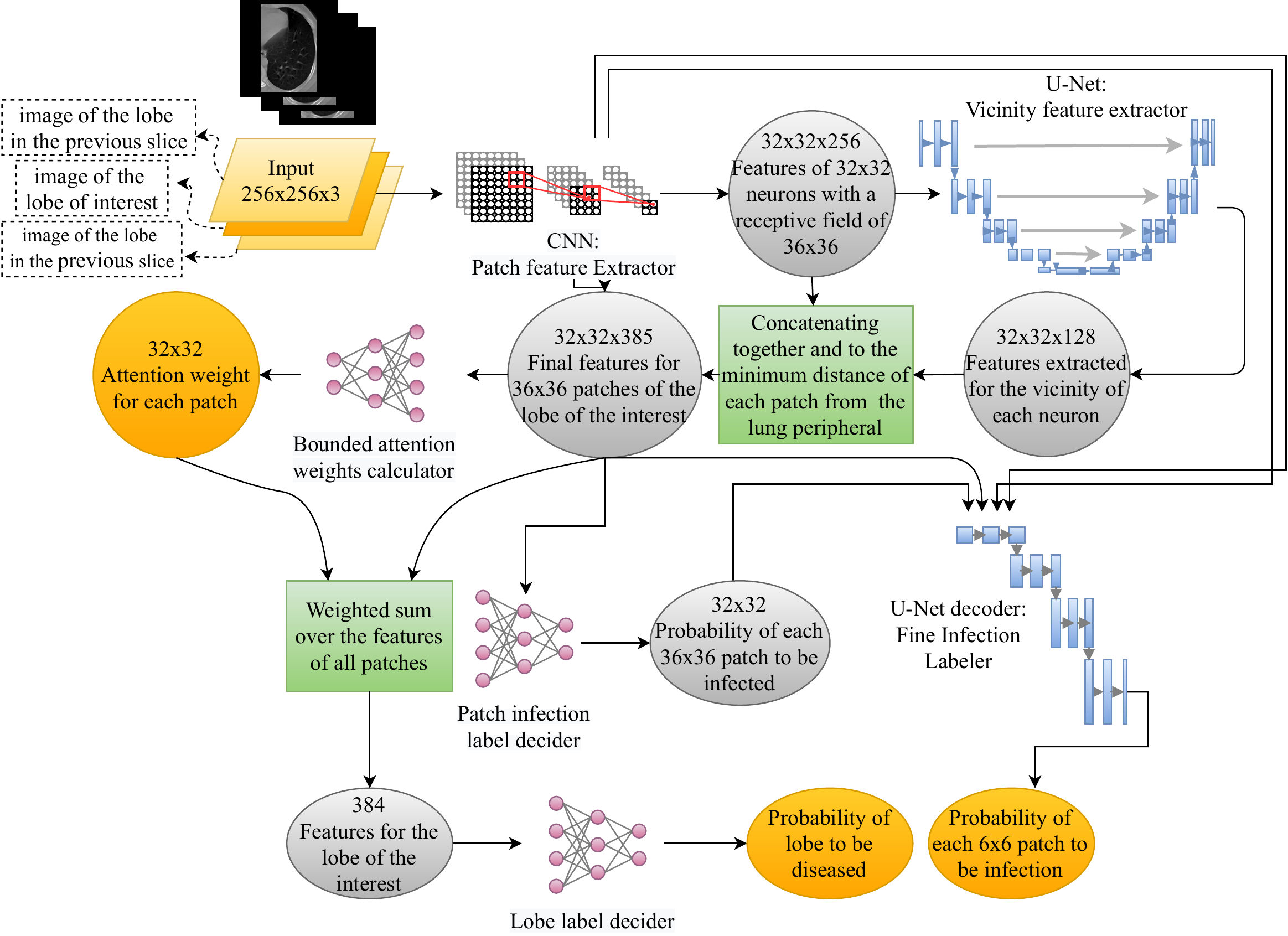}}
\caption{The deep model used for calculating the probability of each slice lobe to be diseased and producing infection map for it.}
\label{fig2}
\end{figure*}

The structure of the proposed deep network that is being used for detecting the lobes having signs of the disease, called the diseased lobes, is presented in Fig. \ref{fig2}. The network receives the images related to one lobe in three consecutive slices as a 256 by 256 by 3 array as input and calculates the features related to the middle slice. The images related to the previous and next slices are used to provide extra information about the continuity of white material of the middle slice. The input is fed into a convolutional subnetwork (Table \ref{tab1}) with an output of size 32 by 32 by 256 which is related to 32 by 32 mesh of neurons with 256 features for each. The receptive field of each neuron in this network is a 36 by 36 patch in the input image so that the extracted features for each neuron are related to the 36 by 36 patch in the input image. 
\par
Making a decision about one patch is not possible by solely using the features extracted from that patch. Therefore, we need to look at a bigger area around the patch in order to predict its state. For example, if a patch is solid white, we don’t know whether the patch is related to the lung periphery or an infected part inside the lung unless we look at its neighboring patches. To add extra information from the vicinity of each patch, the output of the previous subnetwork is fed into a U-Net-style \cite{ronneberger2015u} encoder-decoder (Table \ref{tab1}) so the features are extracted from a bigger receptive field in the encoder part, and the resolution is increased to the input size in the decoder part. In addition to these sets of features, the location of each patch and its distance from the lung peripheral plays an important role in distinguishing COVID-19 from other diseases, as COVID-19 related infections are observed to start from the peripheral. The Manhattan distance between each pixel and the closest peripheral pixel is calculated by using the Breadth-First Search (BFS) algorithm, and the minimum distance of the pixels in each patch is assigned to that patch. The features extracted for each patch, the ones extracted for the vicinity of the patch, and the minimum distance of the patch from the lung peripheral are concatenated to form the final features for each patch. 
\par
Features set of each patch is fed separately into a fully connected subnetwork (Table \ref{tab1}) that calculates the probability of the patch to be infected. Another U-Net style decoder has been used to separate infections in finer details of 6x6 patches from the 36x36 patches that have been identified as infectious. The details of the network's architecture are explained in Table 1. The two-level identification of infections helps in decreasing a large number of false positives by forcing the model to concentrate in larger areas around smaller 6x6 patches. 
\par
The features of all patches are also fed into a fully connected subnetwork (Table \ref{tab1}) to calculate a bounded attention weight between 0 and 1 for each patch. The features of patches are multiplied by the calculated weight and added up to calculate the final features for the lobe. The calculated features are then fed into a fully connected subnetwork (Table \ref{tab1}) that calculates the probability of being infected for the whole slice. 
\par
The mentioned network works only based on patches so the final decision can easily be mapped into the responsible patches. This is a logical scheme for detecting various diseases in the lung as if the network finds any evidence of diseases in a limited number of patches, it will assign it to the entire lobe and in the case of COVID-19, ground-glass opacity infections near the periphery of the lung has been proven to be one of the main pieces of evidence. 

\begin{table}
\caption{Specification of the layers used in each subnetwork of the model used for diseased versus healthy detection}
\label{tab1}
\centering
\setlength{\tabcolsep}{3pt}
\begin{tabular}{|p{100pt}|p{220pt}|}
\hline
Subnetworks & Layers \\\hline

Patch feature extractor & 
conv(64x3x3,s=1), relu, conv(64x3x3,s=1), relu, MaxPool(2x2,s=2), [Name of the output: PFE1]; conv(128x3x3,s=1), relu, conv(128x3x3,s=1), relu, MaxPool(2x2,s=2) [Name of the output: PFE2]; conv(256x3x3,s=1), relu, conv(256x3x3,s=1), relu, MaxPool(2x2,s=2)
\\\hline

Vicinity feature extractor (Encoder) & 
conv(64x1x1,s=1) [Name of the output: VFE1]; conv(64x3x3,s=2), conv(64x3x3,s=1), relu [Name of the output: VFE2]; conv(64x3x3,s=2), conv(64x3x3,s=1), relu
\\\hline

Vicinity feature extractor (Decoder) & 
TransposeConv(64x2x2,s=2), conv(64x3x3,s=1), relu; UNet concatenation with VFE2; TransposeConv(64x2x2,s=2), conv(64x3x3,s=1), relu; UNet concatenation with VFE1  
\\\hline

Patch infection label decider & 
Conv(64x1x1,s=1), relu; Dropout(0.5); Conv(2x1x1,s=1), Softmax
\\\hline

Fine infection label decider & 
Conv(64x1x1,s=1); TransposeConv(64x2x2,s=2), conv(64x3x3,s=1), relu; UNet concatenation with PFE2; TransposeConv(64x2x2,s=2), conv(64x3x3,s=1), relu; UNet concatenation with PFE1; Dropout(0.5), Conv(64x1x1,s=1), relu, Dropout(0.5), Conv(2x1x1,s=1), Softmax
\\\hline

Bounded attention weights calculator & 
Conv(64x1x1,s=1), relu, Dropout(0.5), Conv(2x1x1,s=1), Softmax
\\\hline

Lobe label decider & 
Linear(64), relu, Dropout(0.5), Linear(1), Softmax 
\\\hline

\end{tabular}
\end{table}

\subsection{COVID-19 versus other diseases decision maker}

Another deep network is adopted to distinguish COVID-19 from other diseases. This network also has the same structure as the previous one, except it does not have the part related to detecting infection in the patches. The base part for extracting features for the patches could have been shared between these two networks, but in practice, learning different features for the patches by two separate networks, resulted in a better performance.

\subsection{Aggregating the results of the lobes }

All the lobes from a sample are fed into the aforementioned networks and the probabilities of being diseased versus healthy, and for the disease to be COVID-19 as opposed to other diseases are calculated for each of them. As the signs of diseases should be continuous in the height of the lung, we expect to have at least two consecutive diseased lobes to consider the sample as diseased. Therefore, as shown in (\ref{eq2}), the probability of the left/right lobe of the \textit{i}th slice to be diseased ($P^D_{i,L/R}$) is softened by the probability of the same lobe to be diseased in its adjacent slices, $i - 1$ and $i + 1$. Here, we picked the minimum between the probability assigned to the same lobe and the maximum probability assigned to the two adjacent lobes, as the softened probability of the lobe ($SP^D_{i,L/R}$). Finally, a CT-scan’s probability to be diseased ($P^D$) is calculated as the maximum probability of its slices lobes being diseased (if the probability is more than 0.5, the CT-scan is labeled as diseased). This process eliminates many of the false positives of decision making based on a single slice lobe. The same procedure is also repeated in calculating the probability of a CT-scan to be related to COVID-19 versus other diseases.

\begin{align}
\label{eq2}
\begin{array}{ll}
& SP^D_{i,L/R} = min(P^D_{i-1,L/R}, max(P^D_{i-1,L/R}, P^D_{i+1,L/R})) \\
&\\
& P^D = Max_{i=1}^{n\_slices}(max(SP^D_{i,L}, SP^D_{i,R}))
\end{array}
\end{align}

\subsection{Training the model}

It is typical to use the cross-entropy loss in classification problems, which is in fact maximization of the logarithm of the likelihood. However, in our problem, we have the challenge of having many slices in each sample, and the final decision is made based on the aggregation. Using operations such as maximum in the training phase requires the presence of all images of the slices lobes related to one CT-scan to be in the training batch. Nevertheless, this is not possible in ordinary GPU RAMs as each CT-scan has about 100 slices or 200 lobes, and we also need to have batches of different samples to have more efficient optimization steps. Moreover, this makes the training process time consuming as we need to run a large network over all the images, for every sample. It also has the problem of biasing the gradient flow towards only one or a small number of images. This makes the progress in training slow, and the network needs to be trained for much more epochs to learn something meaningful. It may also result in learning the evidence of the highly infected slices, and lose the power of generalization in slightly infected CT-scans. One solution is to train the network to label the slices lobes instead of the entire CT-scan, but the ground truth labels are not available for slices lobes. We also cannot assign the CT-scan labels to all of their slices lobes as the disease may have affected a part of the lung only. Furthermore, for training the patch labeling part, we require the infected area to be specified in all slices. This is a laborious and time-consuming effort for radiologists, especially at these peak times when thousands of patients are waiting to be examined. Therefore, we utilized a weakly supervised \footnote{Weak supervision is from the perspective that we have sample-level labels but do not have more precise labels for slices and infected zones and they are the ones that are actually needed.} procedure for finding infected slices and infected patches. 

\subsubsection{Weakly supervised training for labeling the samples}

To overcome the problem of unknown labels for slices lobes, a set of 10 nearly equally distanced slices lobes were selected across the height of each sample in the batch. This set of images will somehow cover different parts of the lung, and we can expect that the signs of the disease would be observable in at least some of them. The label of each sample was assigned to 30\% of the top infected slices lobes based on the probabilities assigned by the network, and the cross-entropy loss was calculated only for these top slices lobes. To keep the balance, the top 30\% infected slices lobes were also included in loss calculation of the healthy samples. The loss function for training the diseased versus healthy network samples, is presented in (\ref{eq3}). \textit{D} is 1 for diseased samples and 0 otherwise. \textit{k} is the number of the top diseased slices lobes considered in calculating the loss function and $P^D_{ith\_max}$ is the probability of being diseased assigned by the deep network to the slice lobe having the \textit{ith} maximum probability among all the slices lobes of the sample in the batch. The same procedure was utilized for training the network for detecting COVID-19 vs. other diseases.

\begin{align}
\label{eq3}
%\begin{array}{ll}
loss = \frac{-1}{k} \times \sum_{i=1}^k (&(1 - D) \times ln(1 - p^D_{ith\_max}) + D \times ln(p^D_{ith\_max}))
%\end{array}
\end{align}

\subsubsection{Using fake labels for weakly supervised training of the infections}

To deal with the problem of unknown infection masks for training the patch infection detection module, a semi-infection mask was calculated for each slice lobe. In fact, we highlighted some areas that might have been related to infection in the diseased samples, while all the patches of healthy samples were assumed to have no infection. Therefore, we expected the network to distinguish between the patterns similar to infection that only occur in diseased samples and the ones related to other phenomena (e.g. appearance of the top of the liver), which were common in healthy and diseased samples. In other words, we highlighted the pixels with a color between the lung peripheral and background which included infectious parts and some other parts with similar color (e.g. faded lung peripherals) that were not eliminated by the thresholding process. We used fake infection labels for these areas but in the healthy samples, all of the patches were labeled as healthy.
\par
We took the areas that were neither absolute black nor too white, such as soft tissues and bones, as the suspected areas for infection. To ensure that a pixel was not related to the background or lung's healthy part, we simply needed to use a threshold greater than the standard deviation of all the images in the subject’s CT-scan. In this way, assuming all the images had nearly the same distribution of colors, we could be sure to detect those areas with a confidence level of 95\%. For finding the upper bound that excludes pixels that were too white to be related to infection (tissues and bones), we used the distribution of values in all the images of each subject’s CT-Scan. We expected to see two peaks in the distribution plot, the first one related to the black sides and the second one related to the white sides. We selected the value with the minimum intensity between these two and used it as the threshold to separate them. To assign these pixel-wise infections to the patches of 36x36, a threshold of $\frac{1}{3}$ of density was used. 
\par
Training the infection detection subnetwork was conducted using cross-entropy loss between the predicted probability for the patches and the assigned semi-infection labels. We must note that the proposed labels were not ground truth and might have been wrong. Among those labels, there might have been proposed infected labels that were not related to infection, as well as some missed infected parts. To deal with those wrong labels, we used the difference between 1, as the highest probability, and the probability assigned to the real class of the sample for each patch. Then for the top 80\% patches, the loss was weighted by a factor of 1 and for the rest, it was weighted with a factor of 0.1, so the network could ignore them more easily. As there were many more non-infected patches than the infected ones, the loss related to the infected patches was weighted to keep the balance in the loss injected into the network by both infected and non-infected patches. The weight was calculated according to (\ref{eq4}). In this equation, $N_H$ is the number of patches in healthy samples, $DH$ is the set of all patches of diseased samples labeled as healthy by semi-infection labeler, $DD$ is the set of all patches of diseased samples labeled as diseased, $I$ is the indicator function which is $1$ if its condition is satisfied and $0$ otherwise, and $Top^{0.8}$ stands for being in the top 80\% of patches based on the concordance score.

\begin{equation}
\label{eq4}
w_D = \frac{N_H + \sum_{DH}(I[Top^{0.8}] + 0.1 \times (1 - I[Top^{0.8}])}{\sum_{DD}(I[Top^{0.8}] + 0.1 \times (1 - I[Top^{0.8}])}
\end{equation}

The fine infection labeler network was almost trained in the same manner, with some differences in details. Any 6x6 patch in which more than ¾ of its pixels were labeled as a probable infection, and at least one of the coarser 36x36 patches covering it was identified as infectious with a probability above 0.5, was labeled as infected. The same weighting scheme was adopted to balance the total effect received by the network from both of the non-infected and infected patches. 

\subsubsection{Concordance loss to coordinate the labels of slices and the infection labels of their patches}

For COVID-19, we expect to see the infection in the lobes detected as diseased. Although the patch-labeling part and the lobe-labeling part in the proposed model work on the same set of features, their infection status is decided independently. For example, the lobe-decider may assign the diseased label for one lobe while there is no infected patch with the probability of more than 50\%. To keep the concordance and to make the network more interpretable, an MSE loss was also added between the probability of being diseased assigned by the lobe-labeler and the probability of the most infected coarse patch (36x36). We only added this loss term for healthy and COVID-19 samples as the other diseases might not have included infection. This made the network to bold out at least one infected patch for the slices lobes with high COVID-19 probability and helped it to learn the diseased slices lobes better. 
 
\subsubsection{Parameters of training}
For training the network, we used the total loss which included the loss calculated for labels of the lobes and the concordance loss with a weighting factor of 1, and the loss related to labels of the patches with a weighting factor of 0.01. 10\% of the dataset was used as the test data, 9\% as the validation set to help in hyperparameter and model selection, and the remaining 81\% was used to train the model. The model, except for the fine infection labeler subnetwork, was trained using Adam optimization algorithm \cite{kingma2014adam} with default parameters, and an initial learning rate of 1e-4. Batches of 6 samples with 10 slices for each were chosen to train the network on an Nvidia GForce 2080 Ti GPU with 11 Gb of RAM. The network was trained for 500 epochs and the epoch related to the highest validation accuracy was chosen as the final model. To have a better initial point, the subnetwork for extracting features for the patches was initialized with a VGG-16 network \cite{simonyan2014very} trained on ImageNet \cite{deng2009imagenet} available in the torchvision package \cite{marcel2010torchvision}. After this step of training, the entire network was freezed and only the fine infection labeler network was trained using Adam optimization algorithm with an initial learning rate of 1e-5 using batches of 14 samples with two slices for each sample on a GPU with 6Gb of RAM. 

\subsection{Calculating the percentage of the infected volume of the lung}

As explained in the previous subsections, we detect the lobes of the slices, feed them to a deep model, and the model returns the probability of the patches of each slice that contain the COVID-19 infection. The probability of being infected is calculated for two levels of patches; a coarser infection map for 36x36 patches with an overlap of 28 pixels in each dimension, and a finer infection map for 6x6 patches with an overlap of 4 pixels. For each map, the probability of the patch being infected is assigned to all of its pixels, and for the pixels that are included in multiple patches, the maximum probability is considered. For each map, the pixels having a probability of more than 0.5 are assumed to be infected. The two maps are multiplied with each other to form the final infection map. As a result, the coarser map detects infections in a more robust manner, and the finer map separates the infected area with a higher resolution from the coarse map. The total volume of the infected area is calculated by adding the number of the pixels identified as infected, and the total volume of the lung is calculated by counting the number of the pixels related to the lung lobes in all of the slices which were detected in the preprocessing phase. The ratio of these two numbers determines the percentage of the lung volume that is infected. 

\subsection{Interpreting the decisions made by the model}

As explained in the architecture of the model, the model labels each slice based on an attention mechanism over the features calculated for all 36x36 patches of a slice. In this way, the patches having the marks of diseases that have caused the model to make its decision can be found easily, and verified visually. Moreover, the concordance loss that was defined for making the probability of each slice to have signs of COVID-19 and the probability of its most probable infected coarse patch closer together, would bold out the infected areas, especially the ones that are less visible with the naked eye, at the first glance. 

\section{Results}

\subsection{Data}

We collected 4794 CT-scan samples from Tehran Radiology Center (TRC), Imam Khomeini, Yaas, Rasoul Akram, Firoozgar, and Amiralam Hospitals in Tehran, Afzalipour and Bahonar Hospitals in Kerman, Imam Khomeini Hospital in Qazvin, Isa Ibn Maryam, Amin and Goldis Hospitals in Isfahan, and Shahid Beheshti Hospital in Kashan. The diversity of CT-scan samples has helped us to effectively eliminate batch effects and optimize our system's performance to support images from different CT-scan imaging devices with various properties and cut thicknesses. 
\par
The entire dataset contained 3652 CT-scans from COVID-19 patients approved by RT-PCR tests, 572 healthy CT-scans (from the years before COVID-19 was identified), and 570 CT-scans related to other diseases from before 2019 with similar symptoms as COVID-19. The samples were chosen from patients at different stages of COVID-19, including patients at the initial stage of infection with no visible sign in their CT-scan to the ones in the consolidation stage. The details of the samples are presented in Table \ref{tab2}. The samples from Isa Ibn Maryam, Amin, Goldis, Rasoul Akram, Firoozgar, and Amiralam’s Hospitals were not used in the training phase. Instead, they were utilized to evaluate the performance of the model on CT-scans collected from unseen hospitals. The remaining CT-scan samples were divided into training, validation, and test sets with a ratio of 0.81, 0.09, and 0.1, respectively.

\begin{table}
\caption{The number of samples from different hospitals used for training and evaluation of the model}
\label{tab2}
\centering
\setlength{\tabcolsep}{3pt}
\begin{tabular}{|p{80pt}|p{30pt}|p{50pt}|p{45pt}|p{45pt}|p{55pt}|}
\hline
Hospitals/Centers & Total & COVID-19 patients & Non-Covid-19 patients & Healthy & Status \\\hline

Imam Khomeini (Tehran) & 188 & 188 & 0 & 0 & Used in training \\\hline
Yaas (Tehran) & 38 & 38 & 0 & 0 & Used in training \\\hline
Tehran Radiology Center (TRC) & 438 & 0 & 0 & 438 & Used in training \\\hline
Afzalipour/ Bahonar (Kerman) & 513 & 103 & 276 & 134 & Used in training \\\hline
Imam Khomeini (Qazvin) & 514 & 220 & 294 & 0 & Used in training \\\hline
Shahid Beheshti (Kashan) & 1688 & 1688 & 0 & 0 & Used in training \\\hline

Isa/Amin/Goldis (Isfahan) & 94 & 94 & 0 & 0 & 3 test hospitals \\\hline
Firoozgar (Tehran) & 966 & 966 & 0 & 0 & Test hospital \\\hline
Rasoul Akram (Tehran) & 350 & 350 & 0 & 0 & Test hospital \\\hline
Amiralam (Tehran) & 25 & 25 & 0 & 0 & Test hospital \\\hline

\textbf{Total} & \textbf{4794} & \textbf{3652} & \textbf{570} & \textbf{572} &  \\\hline

\end{tabular}
\end{table}

\subsection{Evaluations}
The model for detecting infected versus healthy samples achieved an accuracy of 96.56\% (sensitivity of 97.25\% for COVID-19, sensitivity of 100\% for other diseases, and specificity of 87.50\% for the healthy samples) over the test data of the seen hospitals, and an average accuracy of 95.8\% for the unseen hospitals. The model for distinguishing COVID-19 from the other diseases achieved an accuracy of 96.12\% (sensitivity of 98.15\% and specificity of 81.03\%) over the test data of the seen hospitals, and 95.73\% for the unseen hospitals. The evaluation metrics calculated for the test data of the hospitals included in the training phase and the unseen hospitals for distinguishing diseased samples from the healthy and COVID-19 samples from the other diseases are presented in Table \ref{tab3}. The evaluation metrics for distinguishing COVID-19 from other diseases are presented with details in Table \ref{tab4}. 

\begin{table}
\caption{Evaluation metrics for the test data from different hospitals for distinguishing diseased samples from healthy samples}
\label{tab3}
\centering
\setlength{\tabcolsep}{3pt}
\begin{tabular}{|p{80pt}|p{45pt}|p{55pt}|p{70pt}|p{50pt}|}
\hline
Hospitals/Centers & Accuracy & Sensitivity (COVID-19) & Sensitivity (Other diseases) & Specificity (Healthy) \\\hline

Tehran Imam Khomeini & 95.24 & 95.24 & - & - \\\hline
Yaas & 100 & 100 & - & - \\\hline
TRC & 95.35 & - & - & 95.35 \\\hline
Kerman & 91.38 & 100 & 100 & 64.29 \\\hline
Qazvin & 98.21 & 96.15 & 100 & - \\\hline
Kashan & 97.02 & 97.02 & - & - \\\hline

Isfahan (unseen) & 93.26 & 93.26 & - & - \\\hline
Firoozgar (unseen) & 96.07 & 96.07 & - & - \\\hline
Rasoul Akram (unseen) & 95.71 & 95.71 & - & - \\\hline
Amiralam (unseen) & 96.00 & 96.00 & - & - \\\hline

\textbf{All (seen)} & \textbf{96.56} & \textbf{97.25} & \textbf{100} & \textbf{87.50} \\\hline
\textbf{All (unseen)} & \textbf{95.80} & \textbf{95.80} & \textbf{-} & \textbf{-} \\\hline

\end{tabular}
\end{table}

\begin{table}
\caption{Evaluation metrics for the test data f different hospitals for distinguishing COVID-19 patients from other diseases}
\label{tab4}
\setlength{\tabcolsep}{3pt}
\centering
\begin{tabular}{|p{80pt}|p{55pt}|p{70pt}|p{75pt}|}
\hline
Hospitals/Centers & Accuracy & Sensitivity (COVID-19) & Specificity (Other diseases) \\\hline

Tehran Imam Khomeini & 100 & 100 & - \\\hline
Yaas & 100 & 100 & - \\\hline
TRC & - & - & - \\\hline
Kerman & 87.18 & 90.91 & 85.71 \\\hline
Qazvin & 76.79 & 76.92 & 76.67 \\\hline
Kashan & 100 & 100 & - \\\hline

Isfahan (unseen) & 88.76 & 88.76 & - \\\hline
Firoozgar (unseen) & 98.65 & 98.65 & - \\\hline
Rasoul Akram (unseen) & 89.14 & 89.14 & - \\\hline
Amiralam (unseen) & 100 & 100 & - \\\hline

\textbf{All (seen)} & \textbf{96.12} & \textbf{98.15} & \textbf{81.03} \\\hline
\textbf{All (unseen)} & \textbf{95.73} & \textbf{95.73} & \textbf{-} \\\hline

\end{tabular}
\end{table}

\begin{figure}[!htb]
\centerline{\includegraphics[width=\columnwidth]{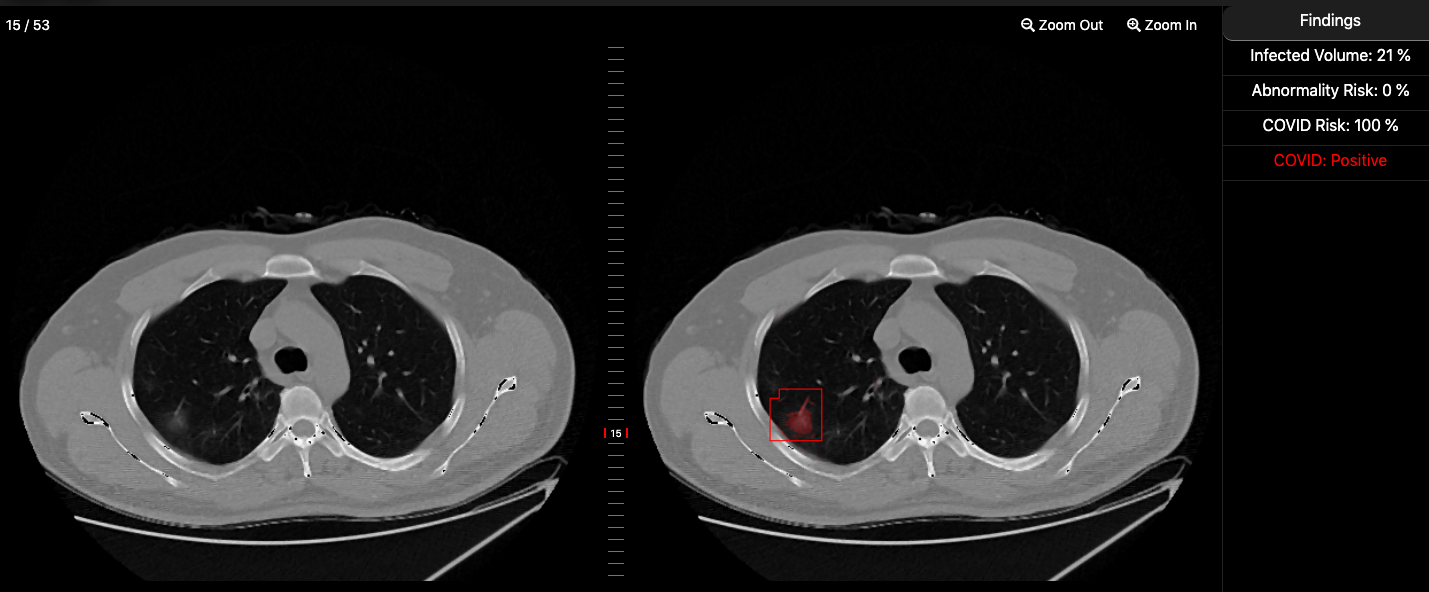}}
\caption{A sample of the proposed system's output. The infected part has been captured and specified with a red box, and the system diagnosis results along with the infected volume are shown on the upper right corner.}
\label{fig3}
\end{figure}

The performance of the model was also evaluated by radiologists in a live session at Tehran Radiology Center (TRC), from which we had only received healthy samples and used them in the training phase. The model labeled all the 6 random samples correctly (1 healthy, 2 COVID-19, and 3 non-COVID diseased samples). Detecting diseased samples while the model was only trained with healthy samples from this center, proves that the model is not biased towards the center’s CT-scan device. The model also detected all the infected areas in the diseased samples, correctly. In addition, we received a report from Imam Ali Hospital in Maragheh that had voluntarily used our system, stating that our algorithm correctly diagnosed COVID-19, healthy, and other diseased samples. An example of the system output is presented in Fig. \ref{fig3}. 

\begin{figure}[!htb]
  \centerline{\includegraphics[width=\columnwidth]{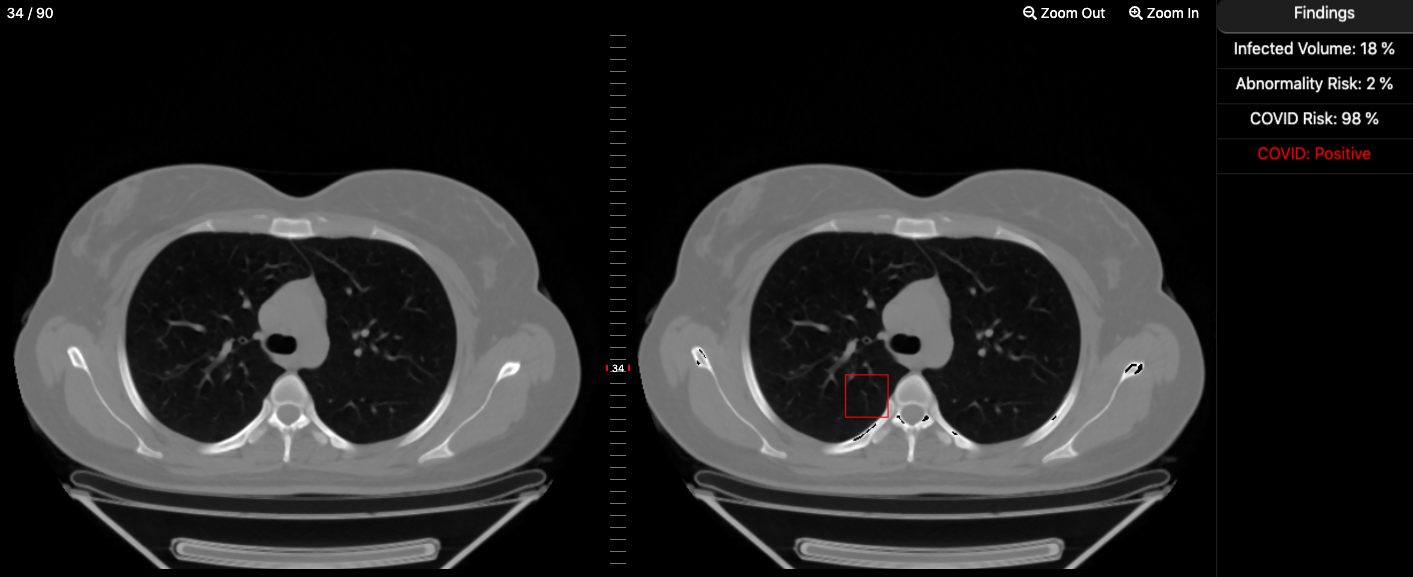}}
  \centerline{\includegraphics[width=\columnwidth]{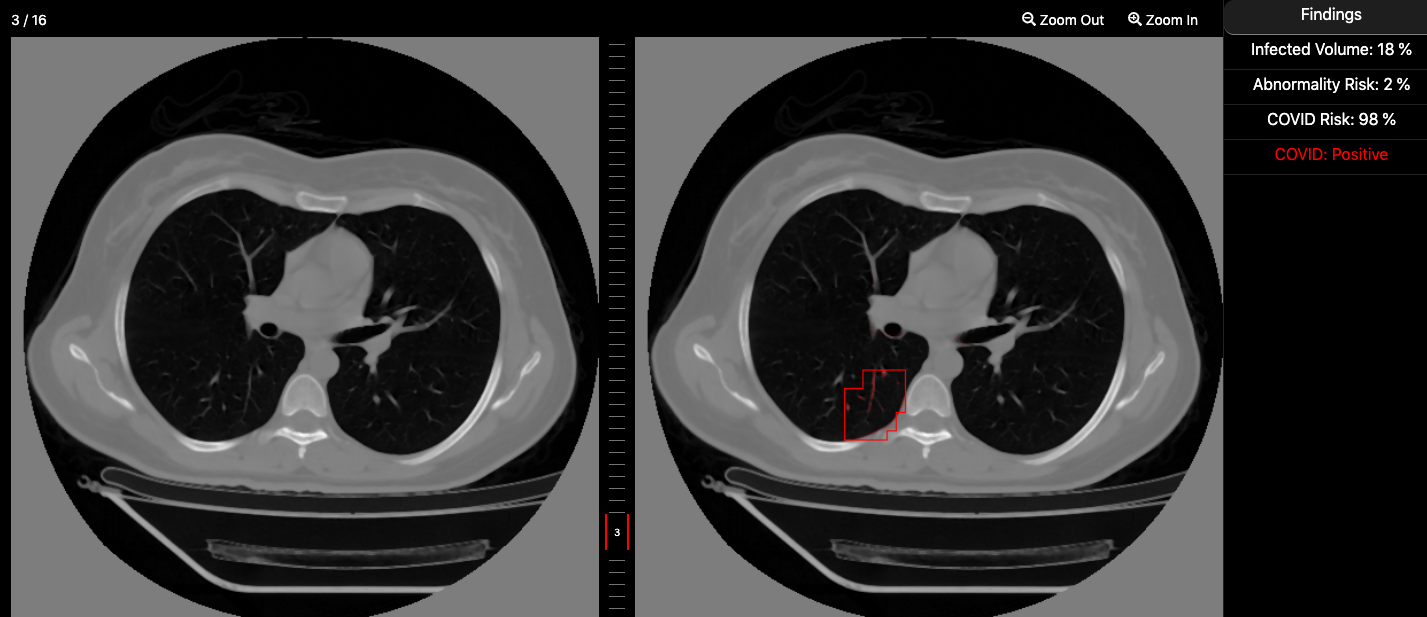}}
\caption{Two slices of two samples with positive RT-PCR that have no obvious visual sign of infection in their CT-Scans. Our system captures a similar pattern in the peripheral sides of the lungs that are near to the spinal cord in most of such samples.}
\label{fig4}
\end{figure}

Our system has also been able to detect COVID-19 from the CT-Scan images with no visual signs of infection. Two such samples are shown in Fig. \ref{fig4}. We have observed that the system captures a similar pattern in the lower parts of the lung which may be a sign that the COVID-19 virus starts from parts closer to the spinal cord and then spreads to the other parts of the lung. 

\section{Conclusion}
In this manuscript, we proposed an interpretable deep learning model for rapid and accurate diagnosis of CT-scan images of lungs to distinguish healthy patients, patients with COVID-19, and patients with other diseases. To address the problem of batch effect in different CT-scan devices and hospitals that is most obvious in the peripheral parts of the lungs, and the background of the images, we separated the lobes from the images and used them as the input to the model. We proposed a unique weakly supervised method for training the model based on the labels of the samples only, without requiring detailed annotations of the slices containing the signs of diseases and the infected area, as it took a great effort to tag all the slices of CT-scan images. Therefore, we were able to use a large cohort for training the model, and hence, making it more generalizable. We gathered 4794 CT-scan images from 12 different hospitals and medical imaging centers of different cities to cover different CT-scan devices, cut thicknesses, and radiation dosages which contained CT-scan images of different stages of COVID-19 along with images from healthy people and patients with pneumonia diseases other than COVID-19 with infections in their lungs. We used samples from half of the centers to train and evaluate the model and another half for evaluating the generalizability of the model on unseen devices. Our trained model reached an accuracy, sensitivity and specificity similar to the top metrics reported in other researchs in distinguishing COVID-19 patients from healthy people and patients with other diseases on a large number of test samples. Reaching a performance similar or better than the models trained in a supervised scheme with fully annotated data shows the efficiency of our unique weakly supervised training method. Furthermore, the model reached a similar accuracy using 1435 samples from the 6 centers not used in the training process, which proves its generalizability. The model also succeeded in detecting diseased and COVID-19 CT-scans from a center used in the training phase from which we only had healthy samples. This proves that the model is not biased toward the batch effect of images related to a special CT-scan device or a specific imaging center. Our model passed the tests successfully in a number of live sessions. We also proved that our model could detect COVID-19 in patients without observable signs of infection in their CT-scans, in the early stages of the disease. 
\par
The high performance, sensitivity, generalizability, and unbiasedness of the proposed model makes it suitable to be used as an AI assistant at medical centers, especially when the pandemic reaches a peak, hospitals are overwhelmed with patients and experience shortage of experts, meaning that expert radiologists may not have the capacity to preform accurate and rapid diagnosis of COVID-19 for all patients.
\par
The system can provide an instant diagnosis on the state of the patients, so they can be quarantined as soon as possible, in order to avoid the spread of virus. The model is also able to label all the slices and show the worst slices to the experienced radiologists for effective diagnosis of the disease. As mentioned before, the model can specify the infected areas of the lung and calculate the total size of the lung so it can calculate the percentage of the infected volume of the lung which can provide a measure for severity of the virus in patients. Therefore, it can be used as a prioritization tool at peak times in order to offer treatment to patients with severe conditions with higher priority. Since the system does not utilize a black-box deep learning model, it can highlight the area of the image that are affected by the virus. This can help radiologists to understand whether the model makes its decisions logically, and use that information to adopt an optimal strategy for treatment of patients. 

\section*{Author Contributions}
HD and HRR designed the experiment. RG, MH, SM, HD, HL, and HRR designed the model and the training method. SM, MH, and HL designed the statistical preprocesses. RG and MH trained the model. RG, MH, SM, AMO, and HL ran the experiments. HG, MAK, BM, MG, OM, OP, KRK, AMA, SH, FAN, AE, MSK, RE, JA, AA, MRS, AK, AF prepared the data, helped the experimental setup and advised the study from the medical point of view. All the authors helped in writing the manuscript. 

\section*{Software availability}
The trained model is provided as a web service and is accessible at https://aimed.ictic.sharif.ir/.

\section*{Acknowledgment}
This work was supported in part by IRI National Science Foundation (INSF) Grant No. 96006077 and ISTI grant number 11/41701. The use of CT-scan images were authorized by Ethical-Code IR.IUMS.REC.1399.008 from Iran University of Medical Sciences.

%\section*{References}

\bibliographystyle{IEEEtran}
\bibliography{main}

\end{document}